# Inevitable Si surface passivation prior to III-V/Si epitaxy:
# A strong impact on wetting properties

S. Pallikkara Chandrasekharan[1], D. Gupta[1], C. Cornet[1,*], and L. Pedesseau[1,*]

[1]*Univ Rennes, INSA Rennes, CNRS, Institut FOTON – UMR 6082, F-35000 Rennes, France*
*Corresponding author: charles.cornet@insa-rennes.fr, laurent.pedesseau@insa-rennes.fr*

Here, we quantitatively estimate the impact of the inevitable Si surface passivation prior to III-V/Si hetero-epitaxy on the surface energy of the Si initial substrate, and explore its consequences for the description of wetting properties. Density Functional Theory is used to determine absolute surface energies of P- and Ga-passivated Si surfaces and their dependencies with the chemical potential. Especially, we show that, while a ≈90 meV/Å² surface energy is usually considered for the nude Si surface, surface passivation by Ga- or P-atoms leads to a strong stabilization of the surface, with a surface energy in the [50-75 meV/Å²] range. The all-*ab initio* analysis of the wetting properties indicate that a complete wetting situation would become possible only if the initial passivated Si surface could be destabilized by at least 15 meV/Å² or if the III-V (001) surface could be stabilized by the same amount.

Heteroepitaxy of III-V semiconductors on silicon has gained renewed interest in the last past years because of its scalability potential for the development of high quality and low-cost devices in the field of photonic or energy applications. While III-V on silicon monolithic integration was recently improved for solar cells, [1] photoelectrochemical cells [2,3] or non-linear photonic devices, [4] the most impressive results were obtained in the field of laser devices developments, where both GaAs-based [5] and GaSb-based [6,7] lasers were epitaxially grown directly on the Si substrate, with performances comparable to state-of-the-art lasers grown on their native substrates. This was made possible thanks to the recent efforts dedicated to the fundamental understanding of III-V/Si heteroepitaxy. [8–16] Importantly, the wetting properties of III-V/Si materials systems and the initial Si surface atomic arrangement were identified as being of central importance for the nucleation of 3D individual islands, and thus subsequently for the generation of various crystal defects, including the so-called antiphase boundaries, [9] with strong consequences on the optoelectronic properties of materials and devices. [3,17,18] Theoretically, the wetting properties are studied within the solid wetting theory, using the Young-Dupré spreading parameter, [19] which definition implies that surface, interface and substrate energies are known.[9] In previous works, complex numerical Density Functional Theory (DFT) developments were used to estimate relative interface energies of III-V/Si interfaces for different atomic scale configurations, [16] which was further extended to the determination of absolute III-V surface and III-V/Si interface energies for different cases. [9,14] In these studies, complete or partial wetting properties were discussed in detail and calculated assuming that the Si substrate remains nude (i.e. without any passivating layer). Although the importance of the Si substrate passivation was pointed out in these previous works, its real impact was not yet quantitatively estimated, leading to a probable overestimation of its value due to a lack of data, such as in ref. [20], where a 92 meV/Å² was chosen for the experimental determination of III-Sb/Si interface energy. In real epitaxial configurations, the Si surface is not expected to remain nude for a long time, as it is a very reactive surface (with a high surface energy). Due to the chemical or thermal preparation of the Si, the surface may contain C, O or H atoms, but it is not expected to be dominant at the surface, otherwise it would create detrimental non-radiative crystal defects. In the ideal situation with optimized conditions, the Si surface is expected to remain quite clean until the III-V growth starts. This is most of the time done by sending group-V or group-III atoms at the surface.

In this paper, we quantitively determine the influence of the passivation by group III or group-V atoms of the initial Si surface on the wetting properties of III-V semiconductors on Si substrates. To this aim, we use DFT to determine the surface energies of Ga-terminated and P-terminated reconstructed Si surfaces over the whole range of chemical potential. From the determination of absolute GaP surface energies, absolute GaP/Si interface energies, and absolute Si-passivated surface energies, we propose a full *ab initio* determination of the spreading parameter for GaP/Si materials. On this quantitative basis, III-V/Si wetting properties are finally discussed.

All the calculations were performed within DFT [21,22] as implemented in the SIESTA [23,24] package with a basis set of finite-range numerical pseudo-atomic orbitals for the valence wave functions. [25] As an exchange-correlation functional, the generalized gradient approximation functional in the PBE [26] form of the Troullier-Martins pseudopotentials were used. [27] The Brillouin zone integration was achieved by 4×4×1 Monkhrost–Pack k-points. [28] A vacuum thickness of 150 Å spacing is introduced in the substrate's vertical direction or in the z-direction to eliminate interactions between the periodic images, following a similar strategy than the one used in previous works. [14]

The perfect nude Si(001) surface whose dangling bonds can disappear by forming dimer bonds to fulfill the Electron

Counting Model [29] (ECM), was described in previous work. [14] Here, we consider the adsorption of foreign atoms on a Si surface that significantly alters its physical and chemical properties. The group-III or group-V passivation of the Si substrate is considered as a perfect monoatomic layer coverage, to mimic the situation commonly reached in epitaxial setups (pressure is set to guarantee such coverage). Note that during the deposition of the first monoatomic layer, the III-V semi-conductor is not formed yet, and thus the physics of the III-V wetting does not apply. Surface reconstructions induced by the insertion of group-III or group-V atoms at the Si(001) surface have already been studied in different works. [30–36]

The P-terminated Si surface is first considered. A nude Si(001) surface consists of Si-Si dimers with one partially occupied orbital per surface atom. In practice, a fully coordinated surface is created when phosphorus interacts with the Si(001) surface. Indeed, the interaction of P with this surface causes the Si-Si dimers to break apart and form P-P dimers instead. [34] Moreover, the extra valence electrons cause the dangling bonds to be replaced with fully occupied orbitals that respect the ECM. Thus, a fully coordinated surface with a 2 x 1 symmetry can be formed, as shown in Fig. 1(a), (b), and (c). Figure 1 illustrates (a) the side view, (b) the top view with the description of a unit cell, and (c) the overall slab model of P passivation on a Si(001) substrate. For the simulation, the slab must be thick enough to mimic the bulk and avoid spurious error. [14] Then, one can have free surfaces at the bottom and material at the top (deposited material). Finally, a sufficient thickness of vacuum is mandatory to avoid any electrostatic charge-charge and dipole-dipole interactions between the slab and its images. [14]

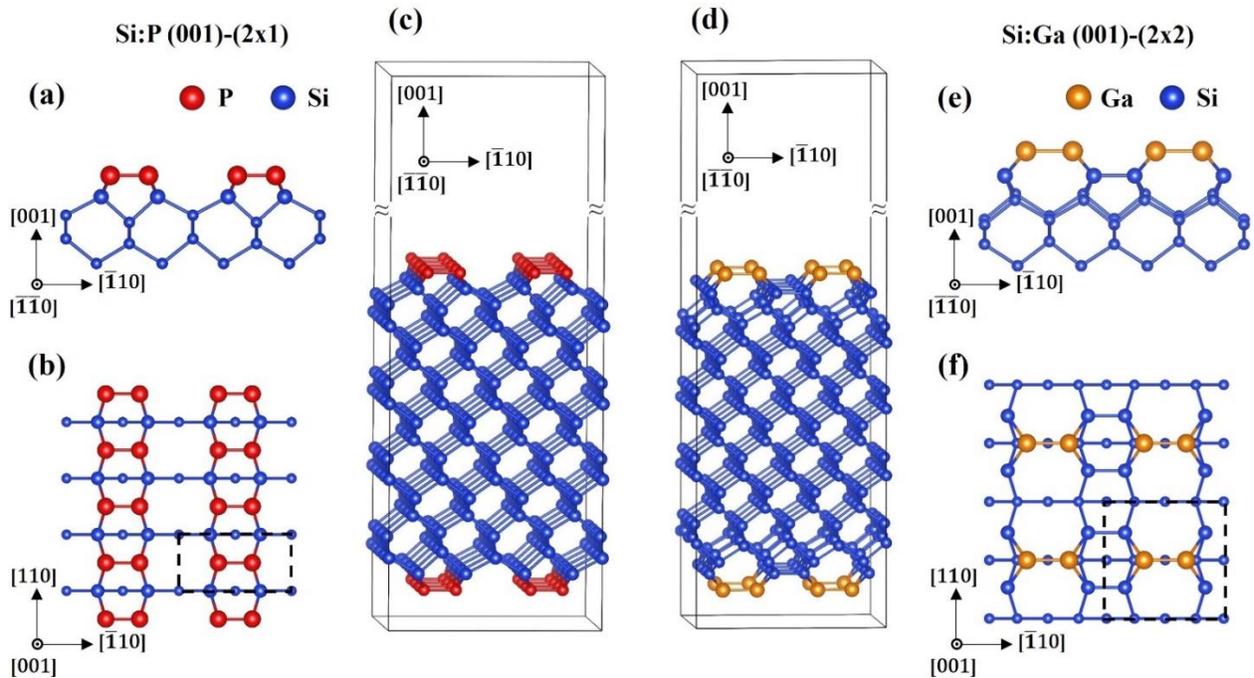

FIG. 1. (a) Side profile, (b) top view, and (c) the slab model for Si:P (001)-(2×1) surface. (d) Side profile, (e) top view, and (f) the slab model for Si:Ga (001)-(2×2) surface. Dashed lines in the top views indicate the unit cells of the reconstructions.

The Ga-terminated Si surface is now studied. Many reconstructed surfaces, including 2x1, 2x2, 2x3, 2x5, and 1x.8, have been experimentally observed for Ga adsorption on Si(001), depending on the adsorption environments. [30–33] We chose a 2x2 para-dimer model for this investigation, which fits experimental findings in several previous works. Fig. 1(d) shows the overall slab model of Ga passivation on Si(001) substrate. In comparison with the Si:P (001)-(2x1) model shown in Fig. 1(a), this 2x2 para-dimer model consists of Ga-Ga dimer aligned above and parallel to the Si-Si dimers, as shown in Fig. 1(e). When Ga dimers positioned between adjacent Si dimer rows, the π-bond of the Si dimer is broken, and Ga-Si bonds are formed. As a consequence, a stable and fully coordinated 2 x 2 symmetry is reached on the top Si surface, as shown in Fig. 1(f). [31] The formation of double bonds in Ga-Ga dimers provides one dangling bond per Ga atom, which violates the ECM and may significantly disturb the growth. [32]

In the following, details of the absolute surface energy calculations for the different surfaces are given. For an accurate determination of surface energies, the same procedure as the one described in ref. [9,14] is used. The absolute surface energy for a non-polar surface is given by:

$$\gamma_{Surface} = \frac{E_{slab} - \sum_i N_i \mu_i}{2A} \quad (1)$$

where $\gamma_{Surface}$ is the surface energy of the non-polar surface, $A$ is the surface area, $E_{slab}$ is the slab energy calculated by DFT (after structural relaxation), $\mu_i$ is the chemical potential of the species $i$ in the slab, and $N$ is the number of particles of the species. Hereafter, we apply the same methodology to determine the absolute surface energy of passivated surfaces. However, the relationship is adapted from equation (1) as follows:

$$\gamma_{Si:X} = \frac{E_{slab} - N_{Si}\mu_{Si-bulk} - N_{SiX}\mu_{SiX-bulk}}{2A} \quad (2)$$

where $\gamma_{Si:X}$ is the absolute surface energy of the passivated surface, $\mu_{Si-bulk}$ is the chemical potential of the Si in its bulk, $N$ is the number of particles of the species, and $\mu_{SiX-bulk}$ is the chemical potential of the thermodynamically stable SiX material. $A$, $E_{slab}$, and $N$ are the parameters described above. It is worth mentioning that in the following, for GaP, we used the heat formation $\Delta H_f$(GaP) equal to -0.928 eV, which defines the length of the interval between the two extreme Ga-rich and P-rich cases. [14] This corresponds to the x axis of Fig. 2.

To determine the surface energy of P-passivated Si surface as a function of the chemical potential variation, we must first define the thermodynamical region where the passivated surface can form and is stable. [37] In fact, from our previous results, we showed that the secondary phase SiP is stable and can be used as a thermodynamic limit before reaching the P-rich region having $\Delta H_f$(SiP) equal to -0.196 eV. [14] On the other side, it can reach the Si-rich region, where the nude Si c(4x2) is stable, as shown in Fig. 2. The grey vertical dotted lines in Fig. 2 corresponds to the threshold crossing point of the chemical potential at which the SiP (to the right) or Si (to the left) can be formed with no cost in energy.

Then, for the surface energy of Ga passivated Si surface, we took Ga-rich and P-rich as the extreme limits of the GaP case, as shown in Fig. 2. Indeed, in this thermodynamic range, no stable secondary phase is reported to our knowledge. As shown in Fig. 2, we highlighted the surface stability regions of the two different passivation strategies studied, by using solid lines.

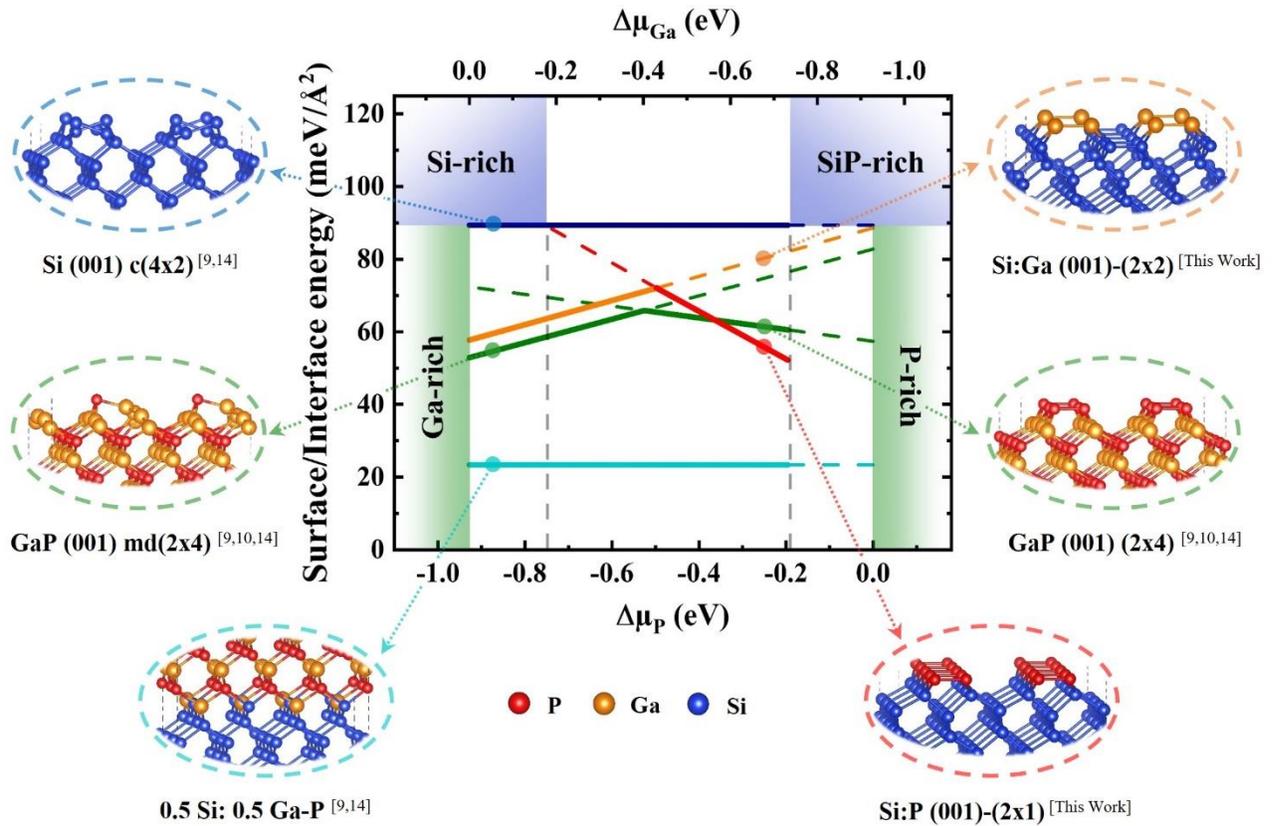

FIG. 2. Absolute surface and interface energy versus $\Delta\mu_P$ and $\Delta\mu_{Ga}$ chemical potentials for the complete GaP/Si system.

In Figure 2, in dark blue, we also reported the most stable nude Si surface, namely the Si(001) c(4x2) surface, whose constant value of the absolute surface energy is 89.37 meV/Å². [9,14] We then also added the two P- or Ga-passivated Si surface energies computed in this work. In red, the absolute surface energy of Si:P (001)-(2x1) is 52.23 meV/Å² (SiP-rich limit) and increases until Si rich conditions. For the passivation of the Si surface with Ga atoms in orange in Fig. 2, the absolute surface energy of Si:Ga (001)-(2x2) varies from 57.74 meV/Å² (Ga-rich limit) to 88.64 meV/Å² ( P-rich limit). Absolute GaP surface energies calculated in ref. [14] are also added in Fig. 2. Specifically, The GaP(001) (2x4) surface whose energy values are 57.4 meV/Å² (P-rich limit) and 72.4 meV/Å² (Ga-rich limit). and the GaP(001) md(2x4) surface whose energy values are 82.8 meV/Å² (P-rich limit) and 52.9 meV/Å²(Ga-rich limit) are drawn. [9,10,14] Finally, in light blue, the most stable absolute GaP/Si interface energy computed in a previous work [14] is also added to Fig. 2. It corresponds to the stable charge-compensated interface, namely the 0.5Si:0.5Ga-P interface, whose value of energy is 23.4 meV/Å². [9,14]

In this section, we investigate the influence of the Si surface passivation on the wetting properties of GaP/Si. To this aim, the absolute energies of the surfaces and interfaces presented above, and their dependencies as a function of the chemical potential are used. The Young-Dupré relation [19] gives the spreading parameter as:

$$\Omega = \gamma^S_{(Si:X)} - \gamma^S_{(GaP)} - \gamma^i_{(GaP/Si)} \qquad (3)$$

where $\gamma^S_{(Si:X)}$ and $\gamma^S_{(GaP)}$ are the absolute surface energies of the most stable passivated silicon and GaP surfaces, which depend on the chemical potential for both. In addition, $\gamma^i_{(GaP/Si)}$ is the absolute interface energy of the most stable GaP/Si interface. The physical meaning of the spreading parameter, $\Omega$, is inherent to its sign. Thus, a positive value of $\Omega$ corresponds to perfect wetting conditions. Conversely, a negative value corresponds to partial wetting, a Volmer-Weber growth. This wetting situation corresponds to the formation of 3D islands whose equilibrium shape depends on $\Omega$.

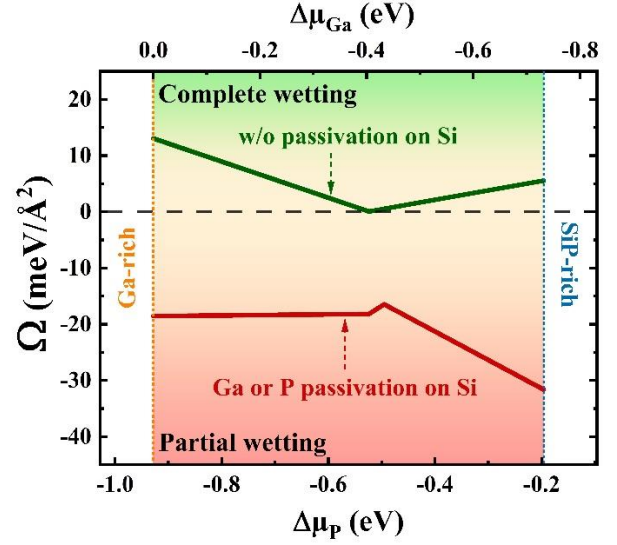

FIG. 3. Young-Dupré spreading parameter $\Omega$ as a function of the chemical potential $\Delta\mu_P$ and $\Delta\mu_{Ga}$, calculated without any passivation (**green**) and with Ga or P passivation (**red**) on the Si surface.

Figure 3 displays the calculated values of $\Omega$ in terms of chemical potential variations of P (bottom axis) and Ga, (top axis). Precisely, we represent the Young-Dupré spreading parameter $\Omega$ without any passivation (in green) and with Ga or P passivation (in red) on the Si surface. The stability domain considered is in between Ga-rich(left limit) and SiP-rich(right limit), respectively. Without Si surface passivation, results from our previous study show a complete wetting scenario, [14] contradicting the experimental observations, [9] as shown by the green solid line. However, with the present more realistic model, the wetting conditions are dramatically reversed by simply including the effect of Ga and P passivation of the Si surface(red) instead of nude Si(green). In details, there are three different behaviors of the red curve (spreading parameter with passivated Si). In the first part ($\Delta\mu_P$ = Ga-rich to -0.523 eV) the spreading parameter remains almost constant, because the variations of Si:Ga (001)-(2x2) and GaP (001) md(2x4) surface energies are very similar, leading to a nearly perfect compensation of both terms (+/-1 meV/Å² variation of $\Omega$ at the maximum) in the spreading parameter. The energy of the interface considered is also a constant and does not depend on the chemical potential, resulting in a spreading parameter stabilized around -18 meV/Å² for this range of chemical potential. In the second narrow region ($\Delta\mu_P$ = -0.523 eV to -0.495 eV), there is a competition between the Si:Ga (001)-(2x2) and GaP (001)-(2x4) surface energies, where the surface energy of the Si:Ga surface is higher than the one of the GaP. As a result of this competition, the spreading parameter $\Omega$ increases with the chemical potential. In the third and final part ($\Delta\mu_P$ = -0.495 eV to SiP-rich), there is a competition between Si:P (001)-(2x1) and GaP (001)-(2x4) surface energies. The energy of the P-passivated Si surface decreases faster than that of the GaP surface, resulting in a general decrease in the spreading parameter Ω towards the SiP-rich limit.

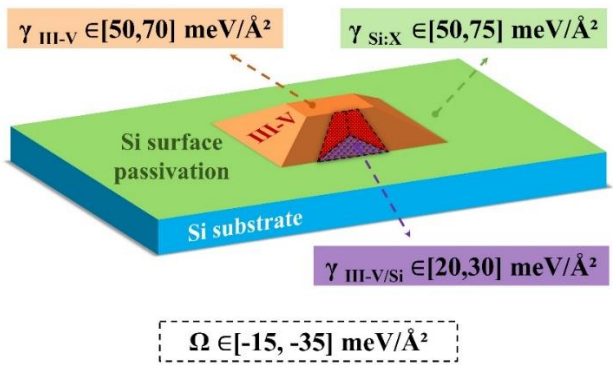

FIG. 4. Scheme summarizing the energies involved in the 3D island formation of GaP on the Si substrate with a range of possible energy values expressed in meV/Å².

From the previous quantitative analysis of surfaces and interfaces energies, and considering the numerical uncertainty that DFT can have for calculating surface and interface energy values in the chemical potential scale, we are now able to provide a first accurate description of the energies involved during III-V/Si heteroepitaxy. As already mentioned in previous studies, although this work is performed on the GaP/Si materials system, the surface and interface energies contributions are likely very similar in most usual Zinc-Blende III-V/Si materials systems. [9,14] Considering the whole range of accessible chemical potentials, III-V surface energies ($\gamma^S_{(III-V)}$) lie in the [50, 70] meV/Å² range, initial Si:X surface energies ($\gamma^S_{(Si:X)}$) lie in the [50, 75] meV/Å² range and III-V/Si interface energies ($\gamma^i_{(III-V/Si)}$), lie in the [20, 30] meV/Å² respectively. All these values suggest that Ω belongs to the [-15, -35] meV/Å² range. Figure 4 summarizes the energies involved for the formation of 3D islands on the passivated Si substrate. The Ω values are negative throughout all the accessible chemical potentials, leading to the formation of 3D islands instead of spreading equally as 2D growth on the Si surface. Overall, a complete wetting situation would become possible (Ω>0) only if the initial Si:X surface could be destabilized by at least 15 meV/Å² or if the III-V (001) surface could be stabilized by the same amount. Reaching a complete III-V/Si wetting during heteroepitaxy would thus certainly imply inserting new atoms in the system, either for stabilizing the III-V surface or for modifying the starting Si surface. The present study gives some quantitative targets to promote complete wetting conditions and reduce coalescence-induced crystal defects.

In conclusion, we have used density functional theory to quantitatively determine the surface energies of Ga- or P-passivated Si surfaces. We show that the inevitable passivation of Si prior to GaP/Si heteroepitaxy leads to a large decreasing of the substrate surface energy, which in turn lead to a large decreasing of the spreading parameter in this system. From this analysis, a synthetic overview of surface and interface energy contributions involved in the wetting properties of III-V/Si heteroepitaxial systems is presented. While the stabilization of the initial Si surface through passivation is expected to be a driving force for 3D Volmer-Weber crystal growth, it is established that a complete wetting situation would become possible if the passivated Si surface could be destabilized by at least 15 meV/Å² or if the III-V (001) surface could be stabilized by the same amount.

This research was supported by the French National Research NUAGES Project (Grant no. ANR-21-CE24-0006). DFT calculations were performed at FOTON Institute OHM – INSA Rennes, and the work was granted access to the HPC resources of TGCC/CINES under the allocation A0140911434 made by GENCI.

[1] M. Feifel et al., *Direct Growth of III–V/Silicon Triple-Junction Solar Cells With 19.7% Efficiency*, IEEE J. Photovolt. **8**, 1590 (2018).

[2] M. Piriyev et al., *Dual Bandgap Operation of a GaAs/Si Photoelectrode*, Sol. Energy Mater. Sol. Cells **251**, 112138 (2023).

[3] L. Chen et al., *Epitaxial III–V/Si Vertical Heterostructures with Hybrid 2D-Semimetal/Semiconductor Ambipolar and Photoactive Properties*, Adv. Sci. **9**, 2101661 (2022).

[4] R. Saleem-Urothodi et al., *Loss Assessment in Random Crystal Polarity Gallium Phosphide Microdisks Grown on Silicon*, Opt. Lett. **45**, 4646 (2020).

[5] S. Chen et al., *Electrically Pumped Continuous-Wave III–V Quantum Dot Lasers on Silicon*, Nat. Photonics **10**, 307 (2016).

[6] M. R. Calvo, L. M. Bartolomé, M. Bahriz, G. Boissier, L. Cerutti, J.-B. Rodriguez, and E. Tournié, *Mid-Infrared Laser Diodes Epitaxially Grown on on-Axis (001) Silicon*, Optica **7**, 263 (2020).

[7] L. Cerutti, D. A. D. Thomas, J.-B. Rodriguez, M. R. Calvo, G. Patriarche, A. N. Baranov, and E. Tournié, *Quantum Well Interband Semiconductor Lasers Highly Tolerant to Dislocations*, Optica **8**, 1397 (2021).

[8] K. Volz, A. Beyer, W. Witte, J. Ohlmann, I. Németh, B. Kunert, and W. Stolz, *GaP-Nucleation on Exact Si (001) Substrates for III/V Device Integration*, J. Cryst. Growth **315**, 37 (2011).

[9] I. Lucci et al., *Universal Description of III-V/Si Epitaxial Growth Processes*, Phys. Rev. Mater. **2**, 060401(R) (2018).

[10] I. Lucci et al., *A Stress-Free and Textured GaP Template on Silicon for Solar Water Splitting*, Adv. Funct. Mater. **28**, 1801585 (2018).

[11] C. Cornet et al., *Zinc-Blende Group III-V/Group IV Epitaxy: Importance of the Miscut*, Phys. Rev. Mater. **4**, 053401 (2020).

[12] M. Rio Calvo, J. Rodriguez, C. Cornet, L. Cerutti, M. Ramonda, A. Trampert, G. Patriarche, and É. Tournié, *Crystal Phase Control during Epitaxial Hybridization of III-V Semiconductors with Silicon*, Adv. Electron. Mater. **8**, 2100777 (2022).

[13] A. Gilbert, M. Ramonda, L. Cerutti, C. Cornet, G. Patriarche, É. Tournié, and J. Rodriguez, *Epitaxial Growth of III-Vs on On-Axis Si: Breaking the Symmetry for Antiphase Domains Control and Burying*, Adv. Opt. Mater. **11**, 2203050 (2023).

[14] S. Pallikkara Chandrasekharan, I. Lucci, D. Gupta, C. Cornet, and L. Pedesseau, *Determination of III-V/Si Absolute Interface Energies: Impact on Wetting Properties*, Phys. Rev. B **108**, 075305 (2023).

[15] O. Romanyuk, T. Hannappel, and F. Grosse, *Atomic and Electronic Structure of GaP/Si(111), GaP/Si(110), and GaP/Si(113) Interfaces and Superlattices Studied by Density Functional Theory*, Phys. Rev. B **88**, 115312 (2013).

[16] O. Supplie, S. Brückner, O. Romanyuk, H. Döscher, C. Höhn, M. M. May, P. Kleinschmidt, F. Grosse, and T. Hannappel, *Atomic Scale Analysis of the GaP/Si(100) Heterointerface by* in Situ *Reflection Anisotropy Spectroscopy and* Ab Initio *Density Functional Theory*, Phys. Rev. B **90**, 235301 (2014).

[17] L. Chen et al., *Strong Electron–Phonon Interaction in 2D Vertical Homovalent III–V Singularities*, ACS Nano **14**, 13127 (2020).

[18] L. Chen, L. Pedesseau, Y. Léger, N. Bertru, J. Even, and C. Cornet, *Antiphase Boundaries in III-V Semiconductors: Atomic Configurations, Band Structures, and Fermi Levels*, Phys. Rev. B **106**, 165310 (2022).

[19] A. DUPRE, *Théorie Mécanique de La Chaleur* (Paris : Gauthier-Villars, 1869).

[20] A. Ponchet, G. Patriarche, J. B. Rodriguez, L. Cerutti, and E. Tournié, *Interface Energy Analysis of III–V Islands on Si (001) in the Volmer-Weber Growth Mode*, Appl. Phys. Lett. **113**, 191601 (2018).

[21] W. Kohn and L. J. Sham, *Self-Consistent Equations Including Exchange and Correlation Effects*, Phys. Rev. **140**, A1133 (1965).

[22] P. Hohenberg and W. Kohn, *Inhomogeneous Electron Gas*, Phys. Rev. **136**, B864 (1964).

[23] E. Artacho et al., *The SIESTA Method; Developments and Applicability*, J. Phys. Condens. Matter **20**, 064208 (2008).

[24] J. M. Soler, E. Artacho, J. D. Gale, A. García, J. Junquera, P. Ordejón, and D. Sánchez-Portal, *The SIESTA Method for* Ab Initio *Order- N Materials Simulation*, J. Phys. Condens. Matter **14**, 2745 (2002).

[25] E. Artacho, D. Sanchez-Portal, P. Ordejon, A. Garcia, and J. M. Soler, *Linear-Scaling Ab-Initio Calculations for Large and Complex Systems*, Phys. Status Solidi B **215**, 809 (1999).

[26] J. P. Perdew, K. Burke, and M. Ernzerhof, *Generalized Gradient Approximation Made Simple*, Phys. Rev. Lett. **77**, 3865 (1996).

[27] N. Troullier and J. L. Martins, *Efficient Pseudopotentials for Plane-Wave Calculations*, Phys. Rev. B **43**, 1993 (1991).

[28] H. J. Monkhorst and J. D. Pack, *Special Points for Brillouin-Zone Integrations*, Phys. Rev. B **13**, 5188 (1976).

[29] M. D. Pashley, *Electron Counting Model and Its Application to Island Structures on Molecular-Beam Epitaxy Grown GaAs(001) and ZnSe(001)*, Phys. Rev. B **40**, 10481 (1989).

[30] J. E. Northrup, M. C. Schabel, C. J. Karlsson, and R. I. G. Uhrberg, *Structure of Low-Coverage Phases of Al, Ga, and In on Si(100)*, Phys. Rev. B **44**, 13799 (1991).

[31] H. Sakama, K. Murakami, K. Nishikata, and A. Kawazu, *Structure of a Si(100)2×2-Ga Surface*, Phys. Rev. B **50**, 14977 (1994).

[32] M. M. R. Evans and J. Nogami, *Indium and Gallium on Si(001): A Closer Look at the Parallel Dimer Structure*, Phys. Rev. B **59**, 7644 (1999).

[33] H. Sakama, A. Kawazu, T. Sueyoshi, T. Sato, and M. Iwatsuki, *Scanning Tunneling Microscopy on Ga/Si(100)*, Phys. Rev. B **54**, 8756 (1996).

[34] Y. Wang, X. Chen, and R. J. Hamers, *Atomic-Resolution Study of Overlayer Formation and Interfacial Mixing in the Interaction of Phosphorus with Si(001)*, Phys. Rev. B **50**, 4534 (1994).

[35] R. I. G. Uhrberg, R. D. Bringans, R. Z. Bachrach, and J. E. Northrup, *Symmetric Arsenic Dimers on the Si(100) Surface*, Phys. Rev. Lett. **56**, 520 (1986).

[36] G. Li and Y.-C. Chang, *Electronic Structures of As/Si(001) 2×1 and Sb/Si(001) 2×1 Surfaces*, Phys. Rev. B **50**, 8675 (1994).

[37] J. E. Northrup, *Structure of Si(100)H: Dependence on the H Chemical Potential*, Phys. Rev. B **44**, 1419 (1991).